\documentclass[aps, prb, twocolumn, showpacs, superscriptaddress]{revtex4}
\usepackage{amsmath, amsfonts, amssymb}
\usepackage{bm}
\usepackage{hyperref}
\usepackage{graphicx}
\usepackage{subfigure}

\begin{document}

\def\Re {\mbox{Re}}
\def\Im {\mbox{Im}}
\newcommand{\avg}[1]{\langle#1\rangle}
\newcommand{\odiff}[2]{\frac{\di #1}{\di #2}}
\newcommand{\pdiff}[2]{\frac{\partial #1}{\partial #2}}
\newcommand{\di}{\mathrm{d}}
\newcommand{\ii}{i}
\newcommand{\norm}[1]{\left\| #1 \right\|}
\renewcommand{\vec}[1]{\mathbf{#1}}
\newcommand{\ket}[1]{|#1\rangle}
\newcommand{\bra}[1]{\langle#1|}
\newcommand{\pd}[2]{\langle#1|#2\rangle}
\newcommand{\tpd}[3]{\langle#1|#2|#3\rangle}
\renewcommand{\vr}{{\vec{r}}}
\newcommand{\vk}{{\mathbf{k}}}
\renewcommand{\ol}[1]{\overline{#1}}

\title{Majorana Edge States in Interacting Two-chain Ladders of Fermions}
\author{Meng Cheng}
\affiliation{Condensed Matter Theory Center, Department of Physics,
             University of Maryland, College Park, Maryland 20742, USA }

\affiliation{Microsoft Research, Station Q, Elings Hall, University of California, Santa Barbara, CA 93106, USA}
\author{Hong-Hao Tu}
\affiliation{Max-Planck-Institut f\"ur Quantenoptik, Hans-Kopfermann-Str. 1, 85748
Garching, Germany}

\date{\today}

\begin{abstract}
	In this work we study interacting spinless fermions on a two-chain ladder with inter-chain pair tunneling while single-particle tunneling is suppressed at low energy. The model embodies a $\mathbb{Z}_2$ symmetry associated with the fermion parity on each chain. We find that when the system is driven to the strong-coupling phase by the pair tunneling, Majorana excitations appear on the boundary. Such Majorana edge states correspond to two-fold degeneracy of ground states distinguished by different fermion parity on each chain, thus representing a generalization of one-dimensional topological superconductors. We also characterize the stability of the ground state degeneracy against local perturbations. Lattice fermion models realizing such effective field theory are discussed.
\end{abstract}
\maketitle

\section{Introduction}
One-dimensional Topological Superconductors (TSC) are novel quantum phases of matter characterized by zero-energy Majorana edge states~\cite{Kitaev_Majorana}. A lot of interest on TSC has been aroused due to the prospect of observing Majorana particles in condensed matter systems~\cite{Wilczek_NP2009} as well as exploiting them as the building blocks of topological quantum computers~\cite{nayak_RevModPhys'08, Alicea_NatPhys2011}. Various proposals of realizing non-Abelian TSC in solid state systems have been put forward, e.g., in semiconductor/superconductor heterostructure~\cite{Sau_PRL10, Lutchyn_PRL2011,Oreg_PRL2010, Alicea_PRB10},  and non-centrosymmetric superconductors~\cite{Sato_PRB09,Sato_PRL2009}.  

The theoretical description of Majorana fermions in TSC is usually based on BCS mean-field Hamiltonian which is essentially a non-interacting theory. Another common feature of most existing studies of one-dimensional TSC is that the BCS pairing (thus the long-range superconducting order) is introduced by proximity effect, since the strong quantum fluctuation in one dimension prevents spontaneous breaking of any continuous symmetry(the Mermin-Wagner theorem). The interplay between Majorana physics and interaction effects remains largely unexplored. Important questions such as how interactions affect Majorana fermions in TSC, and whether one-dimensional TSC can be induced from short-range interactions, have not been fully addressed. Several theoretical studies on the effects of interactions on Majorana fermions in proximity-induced TSC have been performed recently~\cite{Loss_PRL2011, Sela_arxiv, Lutchyn_arxiv_multiband, Stoudenmire_PRB2011}, confirming the stability of Majorana fermions against weak and moderate interactions. On the other hand, it is quite remarkable that the topological classification of one-dimensional non-interacting fermionic systems with time-reversal symmetry is dramatically changed by interactions~\cite{Fidkowski_PRB2011, Fidkowski_PRB2010, Turner_PRB2011}. 

In this work we present a generic field-theoretical model of spinless fermions on two-chain ladders motivated by the second question that whether short-range interactions can induce TSC in one dimension. The model generalizes the simplest one-dimensional TSC, namely spinless fermions with $p$-wave pairing (also known as Majorana chain)~\cite{Kitaev_Majorana}, to interacting two-chain systems. Instead of introducing pairing by proximity effect, the effective field theory includes inter-chain pair tunneling with inter-chain single-particle tunneling being suppressed. Therefore the fermion parity on each chain is conserved. When the pair-tunneling interaction drives the system to strong coupling, localized Majorana zero-energy states are found on the boundaries, which represents a nontrivial many-body collective state of the underlying fermions. We then demonstrate that in a finite-size system the Majorana edge states lead to (nearly) degenerate ground states with different fermion parity on each chain, thus revealing its analogy with the Majorana edge states in non-interacting TSC.  The degeneracy is shown to be robust to any weak intra-chain perturbations, but inter-chain single-particle tunneling and backscattering can possibly lift the degeneracy. We also discuss a lattice model where such field theory is realized at low energy.

\section{Field-Theoretical Model}
We start from an effective field-theoretical description of the model for the purpose of elucidating the nature of the Majorana edge states. We label the two chains by $a=1,2$. The low-energy sector of spinless fermions on each chain is well captured by two chiral Dirac fermions $\hat{\psi}_{L/R, a}(x)$. The non-interacting part of the Hamiltonian is simply given by $\hat{H}_0=\int\di x\,\hat{\mathcal{H}}_0(x)$ where
\begin{equation}
	\hat{\mathcal{H}}_0=-iv_F\sum_{a}\left(\hat{\psi}^\dag_{Ra}\partial_x\hat{\psi}_{Ra}-\hat{\psi}^\dag_{La}\partial_x\hat{\psi}_{La}\right).
	\label{}
\end{equation}

Four-fermion interactions can be categorized as intra-chain and inter-chain interactions. Intra-chain scattering processes (e.g., forward and backward scattering) are incorporated into the Luttinger liquid description of spinless fermions and their effects on the low-energy physics are completely parameterized by the renormalized velocities $v_a$ and the Luttinger parameters $K_a$. We assume that the filling of the system is incommensurate so Umklapp scattering can be neglected. 
For simplicity we assume the two chains are identical so $v_1=v_2=v, K_1=K_2=K$.

We now turn to inter-chain interactions. Those that can be expressed in terms of the densities of the chiral fermions can be absorbed into the Gaussian part of the bosonic theory after a proper change of variables(see below) and we do not get into the details here. We have to consider the pair tunneling and the inter-chain backscattering:
\begin{equation}
	\begin{split}
	\hat{\mathcal{H}}_\text{pair}&=-g_\text{p}(\hat{\psi}_{R2}^\dag\hat{\psi}_{L2}^\dag\hat{\psi}_{L1}\hat{\psi}_{R1}+\text{h.c.})\\
	\hat{\mathcal{H}}_\text{bs}&=g_\text{bs}(\hat{\psi}_{L1}^\dag\hat{\psi}_{R1}\hat{\psi}^\dag_{R2}\hat{\psi}_{L2}+1\leftrightarrow 2).
\end{split}
	\label{eqn:pair}
\end{equation}
The microscopic origin of such terms is highly model-dependent which will be discussed later. The motivation of studying pair tunneling is to ``mimic'' the BCS pairing of spinless fermions without explicitly introducing superconducting pairing order parameter.

The Hamiltonian of the effective theory is then expressed as $\hat{\mathcal{H}}=\hat{\mathcal{H}}_0+\hat{\mathcal{H}}_\text{bs} +\hat{\mathcal{H}}_\text{pair}$. Notice that total fermion number $\hat{N}=\hat{N}_1+\hat{N}_2$ is conserved by the Hamiltonian, but $\hat{N}_1$ and $\hat{N}_2$ themselves fluctuate due to the tunneling of pairs. However, their parities $(-1)^{\hat{N}_a}$ are still separately conserved. Due to the constraint that $(-1)^{\hat{N}_1}\cdot(-1)^{\hat{N}_2}=(-1)^{N}$, we are left with an overall $\mathbb{Z}_2$ symmetry. Therefore we define the fermion parities $\hat{{P}}_a=(-1)^{\hat{N}_a}$, the conservation of which is crucial for establishing the existence and stability of the Majorana edge states and ground state degeneracy. In the following we refer to this overall $\mathbb{Z}_2$ fermion parity as single-chain fermion parity. It is important to notice that the conservation of the single-chain fermion parity relies on the fact that there is no inter-chain single-particle tunneling in our Hamiltonian. We will address how this is possible when turning to the discussion of lattice models.

We use bosonization~\cite{bosonization, delft_review} to study the low-energy physics of the model. The standard Abelian bosonization reads 
\begin{equation}
	\hat{\psi}_{r,a}=\frac{ \hat{\eta}_{r,a}}{\sqrt{2\pi a_0}}e^{i\sqrt{\pi}(\theta_a+r\varphi_a)}
\end{equation}
where $a_0$ is the short-distance cutoff, $r=+/-$ for $R/L$ movers and $\hat{\eta}_{r,a}$ are Majorana operators which keep track of the anti-commuting character of the fermionic operators. We follow the constructive bosonization as being thoroughly reviewed in [\onlinecite{delft_review}]. The two bosonic fields $\varphi_a$ and $\theta_a$ satisfy the canonical commutation relation:
\begin{equation}
  [\partial_x\varphi_a(x), \theta_a(x')]=i\delta(x-x').
  \label{}
\end{equation}
The $\varphi_a$ field is related to the charge density on chain $a$ by $\rho_a=\frac{1}{\sqrt{\pi}}\partial_x\varphi_a$, and $\theta_a$ is its conjugate field, which can be interpreted as the phase of the pair field.

 It is convenient to work in the bonding and anti-bonding basis:
\begin{equation}
\begin{gathered}
  \varphi_{\pm}=\frac{1}{\sqrt{2}}(\varphi_1\pm\varphi_2),\:
\theta_{\pm}=\frac{1}{\sqrt{2}}(\theta_1\pm\theta_2).
\end{gathered}
\end{equation}
The resulting bosonized Hamiltonian decouples as $\hat{\mathcal{H}}=\hat{\mathcal{H}}_++\hat{\mathcal{H}}_-$:
\begin{equation}
  \begin{split}
	  \hat{\mathcal{H}}_+&=\frac{v_+}{2}\left[ K_+(\partial_x\theta_+)^2+{K_+^{-1}}(\partial_x\varphi_+)^2 \right],\\
	  \hat{\mathcal{H}}_-&=\frac{v_-}{2}\left[ K_-(\partial_x\theta_-)^2+{K_-^{-1}}(\partial_x\varphi_-)^2 \right]\\
	  &+\frac{g_\text{p}}{2(\pi a_0)^2 }\cos\sqrt{8\pi}\theta_-+\frac{g_\text{bs}}{2(\pi a_0)^2 }\cos\sqrt{8\pi}\varphi_-.
  \end{split}
  \label{eqn:bosonized}
\end{equation}
Here $a_0$ is the short-distance cutoff. This decoupling of the bonding and the anti-bonding degrees of freedom is analogous to the spin-charge separation of electrons in one dimension.  Without any inter-chain forward scattering, we have
  $K_\pm = K, v_\pm = v$.

  The bonding sector is simply a theory of free bosons. The Hamiltonian in the anti-bonding sector can be analyzed by the perturbative Renormalization Group(RG) method, assuming the bare couplings $g_\text{p}$ and $g_\text{bs}$ are weak. RG flow of the coupling constants are governed by the standard Kosterlitz-Thouless equations~\cite{Jose_PRB1977}: 
\begin{equation}
	\begin{gathered}
		\odiff{y_\text{p}}{l}=(2-2K^{-1}_-)y_\text{p}\\
		\odiff{y_\text{bs}}{l}=(2-2K_-)y_\text{bs}\\
		\odiff{\ln K_-}{l}={2K_-^{-1}}y_-^2,
\end{gathered}
\end{equation}
where $y_-=\frac{g_\text{p}}{\pi v_-}, y_\text{bs}=\frac{g_\text{bs}}{\pi v_-}$ are the dimensionless coupling constants and $l=\ln\frac{a}{a_0}$ is the flow parameter. When $K_->1$(corresponding to attractive intra-chain interaction), $y_\text{p}$ is relevant and flows to strong-coupling under RG flow, indicating gap formation in the anti-bonding sector, while $y_\text{bs}$ is irrelevant so can be neglected when considering long-wavelength, low-energy physics.  Semiclassically, the $\theta_-$ is pinned in the ground state. From now on, we will assume $K_->1$ and neglect the irrelevant coupling $y_\text{bs}$.

\section{Majorana zero-energy edge states}
To clarify the nature of the gapped phase in the anti-bonding sector, we study the model at a special point $K_-=2$, known as the Luther-Emery point~\cite{Luther_PRL1974}, where the sine-Gordon model is equivalent to free massive Dirac fermions. First we rescale the bosonic fields: 
\begin{equation}
  \tilde{\varphi}_-=\frac{\varphi_-}{\sqrt{K_-}},\,\tilde{\theta}_-=\sqrt{K_-}\theta_-,
  \label{}
\end{equation}
and define the chiral fields by $\tilde{\varphi}_{r-}=\frac{1}{2}(\tilde{\varphi}_-+r\tilde{\theta}_-)$. Neglecting the irrelevant backscattering term,
$\hat{\mathcal{H}}_-$ is refermionized to
\begin{equation} \hat{\mathcal{H}}_-=-iv_-(\hat{\chi}_R^\dag\partial_x\hat{\chi}_R-\hat{\chi}_L^\dag\partial_x\hat{\chi}_L)+im(\hat{\chi}_R^\dag\hat{\chi}_L^\dag-\hat{\chi}_L\hat{\chi}_R),
  \label{eqn:refm}
\end{equation}
where the Dirac fermionic fields $\hat{\chi}_r$ are given by 
\begin{equation}
	\hat{\chi}_r=\frac{1}{\sqrt{2\pi a_0}} \hat{\xi}_r e^{ir\sqrt{4\pi }\tilde{\varphi}_{r-}},
	\label{}
\end{equation}
with the fermion mass $m=\frac{g_\text{p}}{\pi a_0}$. $\hat{\xi}_r$ are again Majorana operators. It is quite clear that effective theory \eqref{eqn:refm} also describes the continuum limit of a Majorana chain, which is known to support Majorana edge states~\cite{Kitaev_Majorana}. 

However, caution has to be taken here when dealing with open boundary condition(OBC).  We impose open boundary condition at the level of underlying lattice fermionic operators~\cite{Lecheminant_PRB2002}:
\begin{equation}
	\hat{c}_{ia}\approx\sqrt{a_0}\big[\hat{\psi}_{Ra}(x)e^{ik_Fx}+\hat{\psi}_{La}(x)e^{-ik_Fx}\big],
	\label{}
\end{equation}
where $\hat{c}_{ia}$ are annihilation operators of fermions and $x=ia_0$. Since the chain terminates at $x=0$ and $x=L$, we demand $\hat{c}_0=\hat{c}_{N+1}=0$ where $N=L/a_0$ is the number of sites on each chain. Let us focus on the boundary $x=0$. Thus the chiral fermionic fields have to satisfy $\hat{\psi}_{Ra}(0)=-\hat{\psi}_{La}(0)$.  
Using the bosonization identity, we find $	\varphi_a(0)=\frac{\sqrt{\pi}}{2}$,
from which we can deduce the boundary condition of the anti-bonding field:
\begin{equation}
	\varphi_-(0)=0.
	\label{eqn:phiobc}
\end{equation}

Therefore, we obtain the boundary condition of the Luther-Emery fermionic fields as  $\hat{\chi}_R(0)=\hat{\chi}_L(0)$. The Hamiltonian is quadratic in $\hat{\chi}$ and can be exactly diagonalized by Bogoliubov transformation. We find that the Luther-Emery fields have the following representation:
\begin{equation}
  \begin{split}  
  &\begin{pmatrix}
    \hat{\chi}_R(x)\\
    \hat{\chi}_L(x)
  \end{pmatrix}= \sqrt{\frac{m}{v_-}}\begin{pmatrix}
    1\\
    1
  \end{pmatrix}e^{-mx/v_-}\hat{\gamma}+\dots .
\end{split}
  \label{}
\end{equation}
Here $\dots$ denotes the gapped quasiparticles whose forms are not of any interest to us. The $\hat{\gamma}$ is a Majorana field(i.e., $\hat{\gamma}=\hat{\gamma}^\dag$) and because $[\hat{\mathcal{H}}_-,\hat{\gamma}]=0$, it represents a zero-energy excitation on the boundary.

Now suppose the system has finite size $L\gg \xi=v_-/m$. The same analysis implies that we would find two Majorana fermions localized at $x=0$ and $x=L$ respectively, denoted by $\hat{\gamma}_1$ and $\hat{\gamma}_2$.   As in the case of TSC, the two Majorana modes have to be combined into a (nearly) zero-energy Dirac fermionic mode: $\hat{c}=\frac{1}{\sqrt{2}}(\hat{\gamma}_1+i\hat{\gamma}_2)$. Occupation of this mode gives rise to two degenerate ground states. Tunneling of quasiparticles causes a non-zero splitting of the ground state degeneracy: $\Delta E\approx me^{-L/\xi}$ ~\cite{Cheng_PRL09, Cheng_PRB2010b}.

We notice that very similar technique was previously applied to the spin-$1/2$ edge excitations~\cite{Tsvelik_PRB1990, Lecheminant_PRB2002, Nersesyan_arxiv2011} in the Haldane phase of spin-$1$ Heisenberg chain, the $\mathbb{SO}(n)$ spinor edge states in the $\mathbb{SO}(n)$ spin chain~\cite{Tu_PRL2011} and also the edge state in an attractive one-dimensional electron gas~\cite{Lopatin_unpublished, Seidel_PRB2005}.

To understand the nature of the Majorana edge state, we have to explicitly keep track of the Klein factors which connect states with different fermion numbers. Therefore we separate out the so-called zero mode in the bosonic field ${\phi}_{r,a}$ and write
$\hat{\psi}_{ra}=\frac{1}{\sqrt{2\pi a_0}}\hat{\eta}_{ra}\hat{F}_{ra}e^{ir\sqrt{4\pi}{\phi}_{ra}}$
where the Klein factors $\hat{F}_{ra}$ are bosonic operators that decrease the numbers of $r$-moving fermions on chain $a$ by one~\cite{delft_review}.  Bosonized form of \eqref{eqn:pair} has a product of the Klein factors $\hat{F}_{R2}^\dag\hat{F}_{L2}^\dag \hat{F}_{L1}\hat{F}_{R1}$ in it. Since this term is to be refermionized as $\sim \hat{\chi}_L\hat{\chi}_R$, we are naturally led to define new Klein factors $\hat{F}_r=\hat{F}_{r2}^\dag\hat{F}_{r1}$ for $\hat{\chi}_r$. Notice that so-defined Klein factors satisfy $\{\hat{P}_a, \hat{F}_r\}=0$, {\it i.e.} $\hat{F}_r$ change single-chain fermion parity. Then the fermionic fields that refermionize the sine-Gordon theory at the Luther-Emery point should take the form
\begin{equation}
	\hat{\chi}_r=\frac{1}{\sqrt{2\pi a_0}} \hat{\xi}_r \hat{F}_r e^{ir\sqrt{4\pi}\tilde{\varphi}_{r-}}.
	\label{eqn:chi2}
\end{equation}
Thus one can identify that $\hat{\chi}_r$ corresponds to inter-chain single-particle tunneling. The ground state $\ket{G}$ of the Hamiltonian \eqref{eqn:refm} can be schematically expressed as 
\begin{equation}
	\ket{G}=\exp\left[ \int\di x_1\di x_2\,\hat{\chi}^\dag(x_1)g(x_1,x_2)\hat{\chi}^\dag(x_2) \right]\ket{\text{vac}},
	\label{}
\end{equation}
 where $g(x_1,x_2)$ is the Cooper-pair wave function of the spinless $p$-wave superconductor and $\ket{\text{vac}}$ is the vacuum state of $\hat{\chi}$ fermion. With the definition \eqref{eqn:chi2}, it is easy to check that $\ket{G}$ is a coherent superposition of Fock states having the same single-chain fermion parity, thus an eigenstate of $\hat{{P}}_a$.  On the other hand, the Majorana fermion $\hat{\gamma}$, being a superposition of $\hat{\chi}$ and $\hat{\chi}^\dag$, changes the single-chain fermion parity: $\{\hat{\gamma}, \hat{P}_a\}=0$. As a result, the two degenerate ground states $\ket{G}$ and $\hat{c}^\dag\ket{G}$ have different single-chain fermion parity which is the essence of the Majorana edge states. If the total number of fermions $N$ is even, then the two (nearly) degenerate ground states correspond to even and odd number of fermions on each chain, respectively.

So far all the conclusions are drawn at the Luther-Emery point $K_-=2$. Once we move away from the Luther-Emery point, the theory is no longer equivalent to free massive fermions. An intuitive way to think about the situation is that if we move away from the Luther-Emery point, the $\hat{\chi}$ fermions start to interact with each other. Since the Majorana edge states are protected by the bulk gap as well as the single-chain fermion parity~\cite{Loss_PRL2011, Sela_arxiv, Lutchyn_arxiv_multiband, Stoudenmire_PRB2011}, we expect the qualitative features hold for the whole regime $K_->1$ based on adiabatic continuity.

Notice that the bonding sector remains gapless. In our field-theoretical model, the bonding and anti-bonding degrees of freedom are completely decoupled so the gaplessness of the bonding boson does not affect the degeneracy in the anti-bonding sector.

\begin{figure}
	\begin{center} \includegraphics[width=0.9\columnwidth]{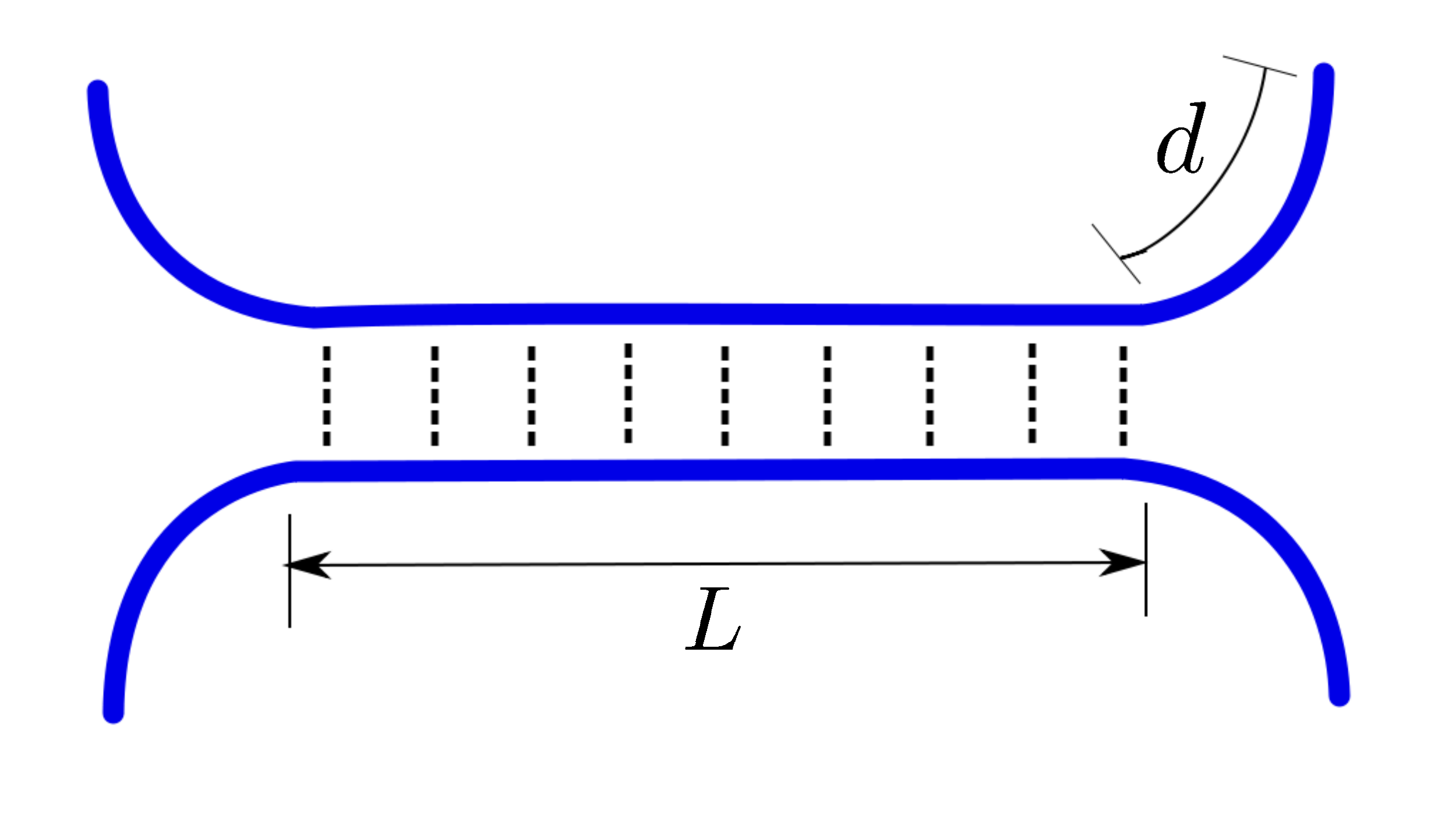} \end{center}
	\caption{Schematic view of the two chains coupled by pair tunneling(denoted by dashed lines). The chains are bended near the two ends to avoid the single-particle tunneling.}
	\label{fig:ladder}
\end{figure}

\section{Stability of the Degeneracy}
We now examine whether the ground state degeneracy we have found has a topological nature. Here we define a topological degeneracy of the ground states by the following criteria: the two degenerate ground states are not distinguishable by any local order parameters(i.e. the difference of the expectation values of any local order parameters in the two ground state must be exponentially small in system size ). By local, we mean local operators in the {\it original} fermionic operators , otherwise we can easily find such an operator in the bosonic representation. For example, in the model \eqref{eqn:bosonized} the operator $\mathcal{O}(x)=\cos\sqrt{2\pi}\theta_-(x)$ can distinguish the two degenerate ground states. But the operator itself is highly non-local in terms of the original fermionic operators.

First of all, by analogy with Majorana chain it is quite obvious that any local operators that involve even numbers of fermion operators on each chain are not able to distinguish the two ground states because such operators always commute with single-chain fermion parity operator. Therefore we only have to consider operators that consist of odd number of single-chain fermion operators. They change the single-chain fermion parity and thus presumably connect the two degenerate ground states. Since all such operators can be decomposed into products of single-particle inter-chain tunneling and backscattering operators, it is sufficient to consider these single-particle operators. 

Let us start with single-particle inter-chain tunneling 
\begin{equation}
\mathcal{O}_\text{T}=\sum_{r=R,L}(\hat{\psi}^\dag_{2r}\hat{\psi}_{1r}+\text{h.c.}).
\end{equation} 
Its bosonic representation is 
\begin{equation}
	\mathcal{O}_\text{T}=\frac{2}{\pi a_0}\cos\sqrt{2\pi}\varphi_-\cos\sqrt{2\pi}\theta_-.
	\label{}
\end{equation}
First let us consider the case when the operator is taken in the bulk of the chain away from any of the boundaries. Because $\theta_-$ is pinned in the ground states, $\varphi_-$ gets totally disordered and therefore $\langle\mathcal{O}_\text{T}\rangle\propto \langle \cos\sqrt{2\pi}\varphi_-\rangle=0$,  which is just equivalent to the fact that the Luther-Emery fermions are gapped. However, this is no longer true as one approaches the ends of the chains, since there exists zero-energy edge states. Let us focus on the left boundary $x=0$. The boundary condition of the anti-bonding boson field $\varphi_-$ has been derived: $\varphi_-(0)=0$. With the boundary condition, we proceed with Luther-Emery solution at $K_-=2$ and find $\mathcal{O}_\text{T}(0)\sim \hat{\chi}(0)+\hat{\chi}^\dag(0)$. Thus $\mathcal{O}_\text{T}(0)$ has nonvanishing matrix element between the two ground states, independent of the system size. As a result, the two-fold degeneracy is splitted.  

We now turn to the inter-chain backscattering 
\begin{equation}
	\begin{split}
	\mathcal{O}_\text{B}&=\hat{\psi}^\dag_{2R}\hat{\psi}_{1L}+\hat{\psi}^\dag_{2L}\hat{\psi}_{1R}+\text{h.c.}\\
	&=\frac{2}{\pi a_0}\cos\sqrt{2\pi}\varphi_+\cos\sqrt{2\pi}\theta_-.
\end{split}
	\label{}
\end{equation}
 An analysis similar to the single-particle tunneling leads to the conclusion that backscattering at the ends also splits the degeneracy. However, even if the backscattering occurs in the middle of the chain, it still causes a splitting of the ground states decaying as a power law in system size $L$. To see this, let us consider a single impurity near the middle of the chain, modeled by $\mathcal{O}_\text{B}(x)$ where $x\approx L/2$. We assume that the backscattering potential is irrelevant under RG flow and study its consequence. The splitting is then proportional to $\langle \cos\sqrt{2\pi}\varphi_+(x)\rangle$ since $\cos\sqrt{2\pi}\theta_-$ has different expectation values on the two ground states. Because $\varphi_+$ is pinned at $x=0$, $\langle \cos\sqrt{2\pi}\varphi_+(x)\rangle\sim 1/x^{K_+}$. Therefore the splitting of the ground states due to a single impurity in the middle of the system scales as $1/L^{K_+}$.

We thereby conclude that the ground state degeneracy is spoiled by the single-particle inter-chain tunneling near the boundaries and the backscattering processes in the bulk. To avoid the unwanted tunneling processes near the ends, one can put strong tunneling barriers between the two chains near the ends, or the chains can be bended outwards so that the two ends are kept far apart~\cite{fisher}, as depicted in Fig. \ref{fig:ladder}.

\section{Lattice Model}
We now show that the field theory \eqref{eqn:bosonized} can be realized in lattice models of fermions. We consider the model of two weakly coupled chains of spinless fermions~\cite{Yakovenko_JETPL1992, Nersesyan_PLA1993, Ledermann_PRB2000, bosonization, Carr_PRB2006}. The Hamiltonian reads
\begin{equation}
	\begin{gathered}
		\hat{H}= -t\sum_{i,a}(\hat{c}_{i+1,a}^\dag \hat{c}_{ia}+\text{h.c.})+\sum_{i,a,r} V(r)\hat{n}_{ia}\hat{n}_{i+r,a}\\
		-t_\perp\sum_i (\hat{c}^\dag_{i2}\hat{c}_{i1}+\text{h.c.}).
	\end{gathered}
	\label{}
\end{equation}
Here $a=1,2$ labels the two chains. We assume the filling is incommensurate to avoid complications from Umklapp scatterings. $V(r)$ is an intra-chain short-range attractive interaction between two fermions at a distance $r$ (in units of lattice spacing). Thus without inter-chain coupling, each chain admits a Luttinger liquid description with two control parameters: charge velocity $v$ and Luttinger parameter $K$ (we assume $V$ is not strong enough to drive the chain to phase separation).

We bosonize the full Hamiltonian and write the theory in the bonding and anti-bonding basis. Hamiltonian in the bonding sector is just a theory of free bosons. In the anti-bonding sector, it reads
\begin{equation}
	\hat{\mathcal{H}}\!=\!\frac{v}{2}\left[K(\partial_x\theta)^2\!+\!\frac{1}{K}(\partial_x\varphi)^2\right]\!+\!\frac{2t_\perp}{\pi a_0}\cos\sqrt{2\pi}\varphi\cos\sqrt{2\pi}\theta.
	\label{eqn:t_perp}
\end{equation}
The bosonic fields $\varphi$ and $\theta$ are in the anti-bonding basis. The perturbation ($t_\perp$) term has {\it nonzero} conformal spin which implies that two-particle processes are automatically generated by RG flow even when they are absent in the bare Hamiltonian. Therefore, one has to include two-particle perturbations in the RG flow
\begin{equation}
\hat{\mathcal{H}}_2=\frac{g_1}{(\pi a_0)^2}\cos\sqrt{8\pi}\varphi+\frac{g_2}{(\pi a_0)^2}\cos\sqrt{8\pi}\theta.
\end{equation}
The RG flow equations for weak couplings have been derived by Yakovenko~[\onlinecite{Yakovenko_JETPL1992}] and Nersesyan {\it et al.}~[\onlinecite{Nersesyan_PLA1993}]. Here we cite their results~\cite{bosonization}:
\begin{equation}
	\begin{split}
		\odiff{z}{l}&=\Big( 2-\frac{K+K^{-1}}{2} \Big)z\\
		\odiff{y_1}{l}&=(2-2K)y_1+(K-K^{-1})z^2\\
		\odiff{y_2}{l}&=(2-2K^{-1})y_2+(K^{-1}-K)z^2\\
		\odiff{ K}{l}&=\frac{1}{2}(y_2^2-y_1^2K^2)
	\end{split},
	\label{eqn:rg2}
\end{equation}
where the dimensionless couplings are defined as $z=\frac{t_\perp a}{2\pi v}$ and $y_{1,2}=\frac{g_{1,2}}{\pi v}$.

Since we are interested in the phase where the pair tunneling dominates at low energy, we assume $K>1$ so $y_1$ is irrelevant and can be put to $0$. Also we neglect renormalization of $K$. Integrating the RG flow equations with initial conditions $z(0)=z_0\ll 1, y_2(0)=0$ we obtain
\begin{equation}
	\begin{split}
		y_2(l)&=z_0^2\frac{K^{-1}-K}{2\alpha}\big[e^{2(1-\alpha)l}-e^{2(1-K^{-1})l}\big]
	\end{split},
	\label{}
\end{equation}
where $\alpha=\frac{1}{2}(K+K^{-1}-2)$. Assume $K^{-1}<\alpha$, then the large-$l$ behavior of $y_2$ is dominated by $e^{2(1-K^{-1})l}$. $y_2$ becomes of order of $1$ at $l^*\approx -\ln z_0/(1-K^{-1})$, where the flow of $z$ yields $z(l^*)\approx z_0^{(\alpha-K^{-1})/(1-K^{-1})}\ll 1$ given $z_0\ll 1$. This means that if $K>\sqrt{2}+1$ (so $K^{-1}<\alpha$), then $y_2$ reaches strong-coupling first. Thus the strong-coupling field theory is given by \eqref{eqn:bosonized}. 

\section{Conclusions}
To conclude, we  consider the strong-coupling phases of model of spinless fermions on a two-chain ladder driven by the pair tunneling. We find that, through Luther-Emery solution of the strong-coupling model, there exists zero-energy excitations on the edges of the ladder represented by Majorana fermions. On a finite system there are always two such Majorana edge states which can be combined to a Dirac fermionic mode and therefore the ground states are two-fold degenerate, corresponding to the mode being occupied or unoccupied. We further clarify the nature of the ground state degeneracy and show that the two states have different fermion parity on each chain. This is in complete analogy with the one-dimensional topological superconductor, where there are two ground states with different {\it total} fermion parity. However, in our case, the one-dimensionality prevents the spontaneous breaking of the global $\mathbb{U}(1)$ symmetry and what we find is the degeneracy between the states with different fermion parity on each chain, subject to the constraint that the total number of fermions is fixed. This is an important distinction between the strong-coupling phase studied in this work and the one-dimensional topological superconductor. What is more, the degeneracy we have found is a purely interaction effect and thus goes beyond the mean-field theory of topological superconductivity in one dimension (essentially non-interacting).

We further characterize the robustness of the ground state degeneracy. We find that the degeneracy is immune to any local perturbations that preserve the single-chain fermion parity. The inter-chain single-particle tunneling in the bulk is prohibited by the existence of a single-particle gap as well. However, near the boundaries the bulk gap vanishes (hence the existence of zero-energy states) and inter-chain single-particle tunneling or backscattering can lift the degeneracy by a finite amount that is independent of the system size. Furthermore, due to the gaplessness of the bonding sector, the inter-chain backscattering in the bulk also change the splitting of the degeneracy to be power-law in system size.

We also discuss a lattice model of two weakly coupled spinless fermion chain where such low-energy effective field theory is realized. We show that there is a range of the Luttinger parameter $K$ such that the inter-chain single-particle tunneling becomes irrelevant (or less relevant) under RG flow, but the two-particle pair tunneling, generated by the single-particle tunneling, becomes relevant and grows to strong coupling eventually. This confirms the validity of our general field-theoretical approach.

{\it Note added}.
During the finalization of the manuscript, we learnt that related works in the context of spin-orbit coupled nanowires has been done by Fidkowski {\it et al.}~\cite{Fidkowski_recent} and also by Sau {\it et. al.}~\cite{Sau_arxiv2011} .

\section{Acknowledgements}
We are grateful to Victor Galitski, Jay D. Sau, Lukasz Fidkowski, Roman Lutchyn, Xiao-Liang Qi and Matthew Fisher for insightful discussions. M.C. is supported by DARPA-QuEST.


\end{document}